\documentclass[10pt,twocolumn]{IEEEtran}
\usepackage[all]{mathtools}
\usepackage[verbose]{cite}
\usepackage[verbose]{wrapfig}
\usepackage[dvipsnames]{xcolor}
\usepackage{cellspace}
\usepackage{moresize}
\usepackage{epsfig,psfrag,subfigure}
\usepackage[font=scriptsize,farskip=5mm]{subfig}
\usepackage{pstricks-add}
\usepackage{amsmath,amssymb}
\usepackage{multirow}
\usepackage{enumerate}
\usepackage{balance}

\begin{document}
\title{\huge{The Capacity of Degraded Cognitive Interference Channel with Unidirectional Destination Cooperation}}

\author{Mohammad~Kazemi,~\IEEEmembership{Student Member,~IEEE}

\thanks{M. Kazemi is with the Department
of Electrical and Computer Engineering, University of Rochester, Rochester,
New York, 14627 USA (e-mail: mkazemi@ece.rochester.edu). }}
 
 \maketitle

\begin{abstract}
Previous works established the capacity region for some special cases of discrete memoryless degraded cognitive interference channel (CIC) with unidirectional destination cooperation (UDC). In this letter, we characterize the capacity region of the general discrete memoryless degraded CIC-UDC. The obtained results imply that the capacity region is achieved by the Gel'fand-Pinsker coding at the cognitive transmitter, superposition coding at the primary transmitter and decode-and-forward at the relay. 
Furthermore, using this general result and a novel converse analysis, we establish the capacity of the Gaussian degraded CIC-UDC, which had been open until this work.
\end{abstract}


\IEEEpeerreviewmaketitle
\vspace{-3mm}
\section{Introduction}

\IEEEPARstart{C}{ognitive} radio (CR) communication has widely considered as a powerful strategy to improve the wireless spectral efficiency. CR networks are subject to interference, which in turn deteriorates data communication rates. A key concern in CR networks, therefore, is how to handle interference and alleviate its destructive effects on the capacity. Cognitive interference channel (CIC) is the setup \cite{c1}, \cite{J1} to study the fundamental limits of communication over CR networks from an information-theoretic perspective. The CIC was further studied in a variety of contexts including Gaussian CIC \cite{c2}, state-dependent CIC \cite{d1,d2}, semi-deterministic CIC \cite{d3}, CIC in better cognitive decoding regime \cite{d3}, less noisy CIC \cite{d4}, and more capable CIC \cite{d4}. A review on the results characterized for the CIC can be found in \cite{d4}.

Cooperative relaying of information is a powerful technique to improve data communication rates \cite{c4}. The CIC with unidirectional destination cooperation (UDC) \cite{a1} - \cite{c3} is a setup to investigate the effect of cooperative relaying on the capacity of CIC. The CIC-UDC, as shown in Fig.~\ref{fig:setup}, is a network with two transmitters communicating two independent messages to two destinations. Communications, however, interfere with each other. Transmitter 1, referred to as the cognitive transmitter, knows both messages 1 and 2, whereas transmitter 2, referred to as the primary transmitter, knows only message 2. Each destination is supposed to decode only its own intended message. Furthermore, destination 1 acts as a relay \cite{c4} and assists destination 2 by sending cooperative information through a relay link. Characterization of the capacity region for the CIC-UDC has remained an open problem. 

The CIC-UDC is defined as degraded if the channel output at destination 2 is degraded with respect to that at destination 1 \cite{c3, c4}. 
The capacity of an special case of the degraded CIC-UDC was previously characterized \cite{c3}, where only destination 1 experiences interference. 
In this letter, we first characterize the capacity region of the \emph{general} discrete memoryless degraded CIC-UDC. Our results show that the Gel'fand-Pinsker coding \cite{c6} at the cognitive transmitter, superposition coding \cite{c6} at the primary transmitter, and decode-and-forward (DF) scheme \cite{c4} at the relay are optimal in the sense of minimizing the effect of interference, thereby achieving the capacity region.
Furthermore, based on this general result and a novel converse analysis, we derive the capacity region of the Gaussian degraded CIC-UDC, which also had been open until this work.
Throughout the letter, random variables (RVs) are indicated by upper case letters. A sequence of RVs $(X_i,X_{i+1},...,Xj)$ is denoted by $X_i^j$. For brevity, $X^j$ is used instead of $X_1^j$.
\vspace{-3mm}
\section{Discrete Memoryless Degraded CIC-UDC}
\subsection{System Model}
\begin{figure}[t]
   \centering
   \includegraphics[scale =.24]{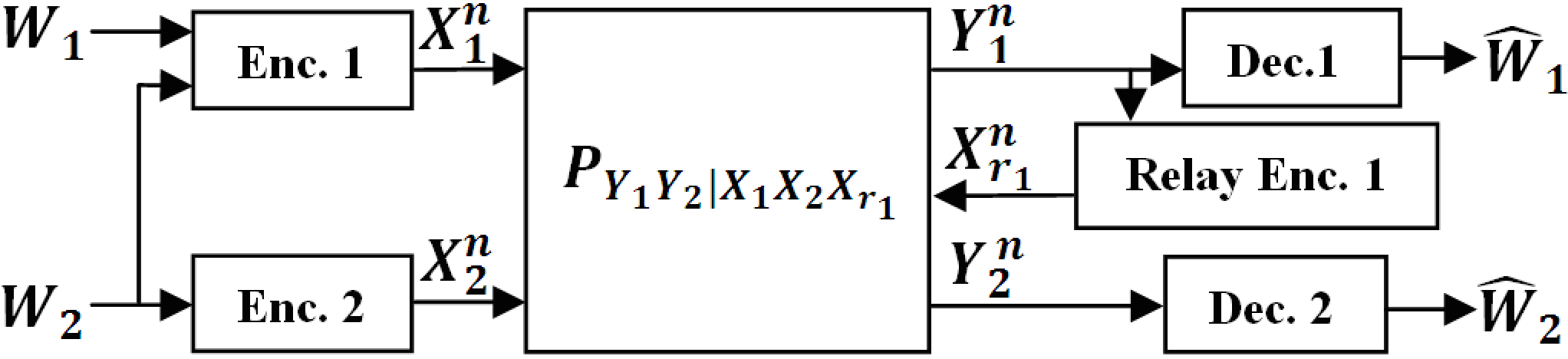} 
   \caption{Discrete memoryless CIC-UDC.}
\label{fig:setup}
\vspace{-.15in}
\end{figure}

Discrete memoryless CIC-UDC, as shown in Fig.~\ref{fig:setup},ß consists of three finite channel input alphabets $\mathcal{X}_1$, $\mathcal{X}_2$, and $\mathcal{X}_{r_1}$, two finite channel output alphabets $\mathcal{Y}_1$ and $\mathcal{Y}_2$, and a set of transition probability distributions $p(y_1,y_2|x_1,x_2,x_{r_1})$. $x_1\in{\mathcal{X}_1}$, $x_2\in{\mathcal{X}_2}$, and $x_{r_1}\in{\mathcal{X}_{r_1}}$ are, respectively, channel inputs from transmitters 1, 2, and relay transmitter. $y_1\in{\mathcal{Y}_1}$ and $y_2\in{\mathcal{Y}_2}$ are, respectively, channel outputs at destination 1 and 2. Channel is degraded if it satisfies the Markov chain $(X_1,X_2)\rightarrow(X_{r_1},Y_1)\rightarrow{Y_2}$ \cite{c3, c6}, thereby
\begin{align}
p(y_1,y_2|x_1,x_2,x_{r_1})= p(y_1|x_1,x_2,x_{r_1})p(y_2|y_1,x_{r_1}).
\label{eq:Deg}
\end{align}
 
\vspace{-5mm}

\subsection{Capacity Region}
\begin{theorem}
For the discrete memoryless degraded CIC-UDC, the capacity region is given by
\begin{align}
&C=\bigcup_{p(u,x_1,x_2,x_{r_1})p(y_1|x_1,x_2,x_{r_1})p(y_2|y_1,x_{r_1})}\nonumber\\
&\left\{
\begin{array}{ll}
&\hspace{-3mm}R_1<I(X_1;Y_1|U,X_2,X_{r_1})\\
&\hspace{-3mm}R_2<\min\{I(U,X_2,X_{r_1};Y_2),I(U,X_2;Y_1|X_{r_1})
\end{array}
\right\}.
\label{eq:Capacity}
\end{align}
\label{thm:CIC-UDC}
\end{theorem}
\vspace{-2mm}
\begin{IEEEproof}
Achievability proof is based on rate splitting and superposition coding \cite{c6} at Transmitter 1, and DF scheme \cite{c4} at the relay. Regular encoding and slide window decoding \cite{c6} is adopted for the DF scheme. Transmitter 2 encodes its message independently. Details of the achievability proof are similar to those of the proof of [13, Lemma 1] except that here destination 1 is supposed to decode only message $W_1$, but in [13, Lemma 1], destination 1 needs to decode both messages $W_1$ and $W_2$.
Converse proof is as follows. Define auxiliary random variable $U_i$, $i\in\{1,...,n\}$, such that $U_i = (W_2,Y_1^{i-1}, Y_2^{i-1})$. From Fano's inequality, we consider
\begin{align}
&\hspace{-2mm}nR_1-n\delta_{1,n}=H(W_1)-n\delta_{1,n}\leq{I(W_1;Y_1^n|W_2)}\label{1}\\
&\hspace{-2mm}\leq{I(W_1;Y_1^n,Y_2^n|W_2)}\nonumber\\
&\hspace{-2mm}=\sum_{i=1}^n{I(W_1;Y_{1,i},Y_{2,i}|W_2,Y_1^{i-1},Y_2^{i-1})}\label{2}\\
&\hspace{-2mm}=\sum_{i=1}^n{I(W_1,X_{1,i};Y_{1,i},Y_{2,i}|W_2,Y_1^{i-1},Y_2^{i-1},X_{2,i},X_{r_1,i})}\label{3}\\
&\hspace{-2mm}=\sum_{i=1}^n{I(W_1,X_{1,i};Y_{1,i}|W_2,Y_1^{i-1},Y_2^{i-1},X_{2,i},X_{r_1,i})}\nonumber\\
&\hspace{-2mm}+\sum_{i=1}^n{I(W_1,X_{1,i};Y_{2,i}|W_2,Y_1^{i-1},Y_2^{i-1},X_{2,i},X_{r_1,i},Y_{1,i})}\label{4}\\
&\hspace{-2mm}=\sum_{i=1}^n{I(W_1,X_{1,i};Y_{1,i}|U_i,X_{2,i},X_{r_1,i})},\label{R1}
\end{align}
\begin{equation}
\hspace{-2mm}=\sum_{i=1}^n{I(X_{1,i};Y_{1,i}|U_i,X_{2,i},X_{r_1,i})},\label{R1f}
\end{equation}
where~\eqref{1} follows as $W_1$ and $W_2$ are independent,~\eqref{2} follows from the chain rule,~\eqref{3} follows as $X_{1,i}$, $X_{2,i}$ and $X_{r_1,i}$ are, respectively, deterministic functions of $(W_1,W_2)$, $W_2$, and $Y_1^{i-1}$. As the channel is memoryless and satisfies the degradedness condition~\eqref{eq:Deg}, the second term of~\eqref{4} is equal to zero. Hence, equality~\eqref{R1} is established. Finally,~\eqref{R1f} follows from Markov chain $(W_1,U_i)\rightarrow(X_{1,i},X_{2,i},X_{r_1,i})\rightarrow{Y_{1,i}}$.
To obtain an upper bound on $R_2$, we consider
\begin{align}
n&R_2-n\delta_{2,n}=H(W_2)-n\delta_{2,n}\leq{I(W_2;Y_2^n)}\nonumber\\
&\leq{I(W_2;Y_1^n,Y_2^n)}=\sum_{i=1}^n{I(W_2;Y_{1,i},Y_{2,i}|Y_{1}^{i-1},Y_{2}^{i-1},X_{r_1,i})}\nonumber\\
&=\sum_{i=1}^n{I(W_2;Y_{1,i}|Y_{1}^{i-1},Y_{2}^{i-1},X_{r_1,i})}\nonumber\\&\hspace{11mm}+\sum{I(W_2;Y_{2,i}|Y_{1}^{i-1},Y_{2}^{i-1},Y_{1,i},X_{r_1,i})}\nonumber\\
&=\sum_{i=1}^n{I(W_2;Y_{1,i}|Y_{1}^{i-1},Y_{2}^{i-1},X_{r_1,i})}\label{R2Deg}\\
&\leq{\sum_{i=1}^n{H(Y_{1,i}|X_{r_1,i})-H(Y_{1,i}|Y_1^{i-1},Y_{2}^{i-1},W_2,X_{r_1,i},X_{2,i})}}\nonumber
\end{align}
\begin{equation}
=\sum_{i=1}^n{I(U_i,X_{2,i};Y_{1,i}|X_{r_1,i})}\label{R2G},
\end{equation}
where~\eqref{R2Deg} follows since the channel is memoryless and satisfies the degradedness condition (1).
The second upper bound on $R_2$ is established as follows,
\begin{align}
&\hspace{-2mm}nR_2-n\delta_{2,n}=H(W_2)-n\delta_{2,n}\leq{I(W_2;Y_2^n)}\nonumber\\
&\hspace{-2mm}=\sum_{i=1}^n{I(W_2;Y_{2,i}|Y_2^{i-1})}\nonumber\\
&\hspace{-2mm}\leq\sum_{i=1}^n[H(Y_{2,i})-H(Y_{2,i}|Y_1^{i-1},Y_2^{i-1},W_2,X_{2,i},X_{r_1,i})]\label{R22}
\end{align}
\begin{equation}
\hspace{-2mm}=\sum_{i=1}^n{I(U_i,X_{2,i},X_{r_1,i};Y_{2,i})},\label{R2G1}
\end{equation}
where~\eqref{R22} follows from the fact that conditioning does not increase entropy.
Note that $\delta_{1,n}$ and $\delta_{2,n}$ goes to zero as $n$ goes to infinity. The rest of the proof is straightforward by introducing an independent time sharing RV $Q$, uniformly distributed over $[1,n]$, and defining $U = (U_Q,Q)$, $X_1=X_{1,Q}$, $X_2=X_{2,Q}$, $X_{r_1}=X_{r_1,Q}$, $Y_1=Y_{1,Q}$, and $Y_2=Y_{2,Q}$. 
\end{IEEEproof}

\vspace{-1mm}
\section{Gaussian Degraded CIC-UDC}

\subsection{System Model}
Let $\{Z_{1,i}\}_{i=1}^n$ and $\{Z_{2,i}\}_{i=1}^n$ be independent sequences of i.i.d. zero-mean Gaussian random variables with variance $N_1$ and $N_2$, respectively. At the $i$th transmission, $X_{1,i}$, $X_{2,i}$, and $X_{r_1,i}$ are sent, and
\begin{subequations}
\begin{align}
Y_{1,i}&=X_{1,i}+aX_{2,i}+Z_{1,i}\label{GC1}\\
Y_{2,i}&=Y_{1,i}+X_{r_1,i}+Z_{2,i},\label{GC2}
\end{align}
\end{subequations}
are received, which satisfy the degradedness condition represented by {\eqref{eq:Deg}}. Parameter $a$ is a real constant, and channel inputs are subject to average power constraints, i.e., 
\begin{equation}
\sum_{i=1}^n{X_{1,i}^2}\leq{nP_1},~
\sum_{i=1}^n{X_{2,i}^2}\leq{nP_2},~
\sum_{i=1}^n{X_{r_1,i}^2}\leq{nP_{r_1}}.
\label{PowerConst}
\end{equation}
\begin{figure*}
\normalsize
\begin{align}
C=\bigcup_{\beta,\gamma}
\left\{
\begin{array}{l}
(R_1,R_2):
R_1<\Psi{\left(\dfrac{(1-\gamma^2)P_1}{N_1}\right)}\\
R_2<\max_{\alpha}\min\left\{
\begin{array}{ll}
\Psi{\left(\dfrac{(\gamma^2P_1(1-\bar{\alpha}\beta)+a^2{\alpha}P_2+{2a\alpha\gamma}\sqrt{{\beta}P_1P_2}}{(1-\gamma^2)P_1+N_1}\right)}\\
\Psi{\left(\dfrac{\gamma^2P_1+a^2P_2+P_{r_1}+2a\gamma\sqrt{{\beta}P_1P_2}+2a\sqrt{\bar{\alpha}P_{r_1}P_2}+2\gamma\sqrt{\bar{\alpha}{\beta}P_{r_1}P_1}}{(1-\gamma^2)P_1+N_1+N_2}\right)}
\end{array}
\right\}
\end{array}
\right\}
\label{GaussianCapacity}
\end{align}
\hrulefill
\end{figure*}
\vspace{-.15in}
\subsection{Capacity Region}
\emph{Theorem 2:} For the Gaussian CIC-UDC, the capacity region  is given by~\eqref{GaussianCapacity}, at the top of the next page, where $\alpha\in{[0,1]}$, $\beta\in{[0,1]}$, $\gamma\in{[-1,1]}$, $\bar{\alpha}\triangleq{1-\alpha}$, and $\Psi(x)\triangleq\dfrac{1}{2}\log(1+x)$.

\vspace{2mm}
\emph{Proof:} Achievability proof follows from {\eqref{eq:Capacity}} by computing the mutual information terms with $(U,X_1,X_2,X_{r_1})$ having the following Gaussian distributions: $X_{r_1}\sim\mathcal{N}(0,P_{r_1})$, $X_2=\sqrt{{\bar{\alpha}P_2}/{P_{r_1}}}X_{r_1}+X_2'$, $U=\sqrt{{\beta{P_1}}/{P_2}}X_2+U'$, and $X_1=U+X_1'$.
Here, $X_{r_1}$, $X_2'\sim\mathcal{N}(0,\alpha{P_2})$, $U'\sim\mathcal{N}(0,\gamma^2\bar{\beta}P_1)$, and $X_1'\sim\mathcal{N}(0,(1-\gamma^2)P_1)$ are independent Gaussian RVs. 

\vspace{2mm}
Before delving into the converse proof, we need to state some properties of Gaussian RVs. 

\vspace{1mm}
\begin{lemma}
For jointly Gaussian random variables $\{V_{1,i},V_{2,i}\}_{i=1}^n$, we have:
\begin{subequations}
\begin{align}
\left|\sum_{i=1}^{n}{E[V_{1,i}V_{2,i}]}\right|&\leq{\sqrt{K\times{L}}}\label{L11}\\
{\sum_{i=1}^{n}{h(V_{1,i}|V_{2,i})}}&\leq{\frac{1}{2}\log2\pi{e}\left(K-M\right)}\label{L12}.
\end{align}
\end{subequations}
where $K=\sum_{i=1}^{n}E[{V_{1,i}^2}]$, $L=\sum_{i=1}^{n}{E\left[E^2[V_{2,i}|V_{1,i}]\right]}$, and $M=\sum_{i=1}^{n}{E\left[E^2[V_{1,i}|V_{2,i}]\right]}$.
\label{lem:Gauss}
\end{lemma}

\vspace{2mm}
\begin{IEEEproof}
Proof is detailed in Appendix A.
\end{IEEEproof}

\vspace{1mm}
We now present converse proof of Theorem 2.
First, we establish an upper bound on $R_1$. Following from~\eqref{R1f}, we have
\begin{align}
&R_1-\delta_{1,n}\leq \dfrac{1}{n}\sum_{i=1}^{n}{I(X_{1,i};Y_{1,i}|U_i,X_{2,i},X_{r_1,i})}\nonumber\\
&=\dfrac{1}{n}\sum_{i=1}^{n}{h(Y_{1,i}|U_i,X_{2,i},X_{r_1,i})}-\frac{1}{2}\log2\pi{e}N_1\nonumber\\
&=\dfrac{1}{2}\log2\pi{e}((1-\gamma^2)P_1+N_1)-\frac{1}{2}\log2\pi{e}N_1\label{rg1}\\
&=\Psi\left(\dfrac{(1-\gamma^2)P_1}{N_1}\right)\label{rg2}.
\end{align}
where~\eqref{rg1} follows from Lemma~\ref{lem:h1} below.

\vspace{1mm}
\begin{lemma}
Let ${h_1\triangleq \frac{1}{n}\sum_{i=1}^{n}{h(Y_{1,i}|U_i,X_{2,i},X_{r_1,i})}}$. Then, 
\begin{equation}
h_1=\dfrac{1}{2}\log2\pi{e}((1-\gamma^2)P_1+N_1)\label{A},
\end{equation}
where $\gamma\in[-1,1]$.
\label{lem:h1}
\end{lemma}

\vspace{1mm}
\begin{IEEEproof}
$h_1$ can be lower bounded as
\begin{align}
h_1&\geq{\dfrac{1}{n}\sum_{i=1}^{n}{h(Y_{1,i}|U_i,X_{1,i},X_{2,i},X_{r_1,i})}}\nonumber\\
&=\dfrac{1}{n}\sum_{i=1}^{n}{h(Y_{1,i}|X_{1,i},X_{2,i},X_{r_1,i})}
=\dfrac{1}{2}\log{2\pi{e}N_1}.\label{A1}
\end{align}
On the other hand, using the definition of $Y_{1,i}$ in~\eqref{GC1}, $h_1$ can be upper bounded as
\begin{align}
h_1&=\dfrac{1}{n}\sum_{i=1}^{n}{h(X_{1,i}+aX_{2,i}+Z_{1,i}|U_i,X_{2,i},X_{r_1,i})}\nonumber\\
&=\dfrac{1}{n}\sum_{i=1}^{n}{h(X_{1,i}+Z_{1,i}|U_i,X_{2,i},X_{r_1,i})}\nonumber\\
&\leq{\dfrac{1}{n}\sum_{i=1}^{n}{h(X_{1,i}+Z_{1,i}})}\nonumber\\
&\leq{\dfrac{1}{2}\log{2\pi{e}}\left(\dfrac{1}{n}\sum_{i=1}^{n}{E[X_{1,i}^2+Z_{1,i}^2]}\right)}\label{A21}\\
&\leq\dfrac{1}{2}\log2\pi{e}(P_1+N_1)\label{A22},
\end{align}
where~\eqref{A21} follows from~\eqref{L12}, and~\eqref{A22} follows from~\eqref{PowerConst}.
Comparing lower bound~\eqref{A1} and upper bound~\eqref{A22}, there exists some $\gamma\in[0,1]$ such that~\eqref{A} is concluded.
\end{IEEEproof}

Next, we  establish a few properties of the random sequences $X_1^n$, $X_2^n$, and $X_{r_1}^n$ that will be used in the sequel.

\vspace{2mm}
\begin{lemma}
Define $S_1=\frac{1}{n}\sum_{i=1}^{n}{E\left[E^2[X_{1,i}|X_{r_1,i}]\right]}$, $S_2=\frac{1}{n}\sum_{i=1}^{n}{E\left[E^2[X_{1,i}|X_{2,i}]\right]}$, $S_3 = \frac{1}{n}\sum_{i=1}^{n}E[X_{1,i}X_{2,i}]$, $S_4 = \frac{1}{n}\sum_{i=1}^{n}{E[X_{r_1,i}(X_{1,i}+aX_{2,i})]}$, and $S_5 = \frac{1}{n}\sum_{i=1}^{n}{E\left[E^2[X_{1,i}+aX_{2,i}|X_{r_1,i}]\right]}$. Then, we have
\vspace{2mm}
\begin{enumerate}[(a)]
	\item $\max\{S_1,S_2\}=\beta\gamma^2{P_1}$ \vspace{1mm}\label{L21}
	\item $S_3\leq{\sqrt{\gamma^2\beta{P_1P_2}}}$, \vspace{1mm}\label{L22}
	\item $|S_4|=\sqrt{P_{r_1}}\left(a\sqrt{\bar{\alpha}P_2}+\sqrt{\gamma^2{\beta}{\bar{\alpha}}{P_1}}\right)$,\label{L23}
	\item $S_5\geq{\left(a\sqrt{\bar{\alpha}P_2}+\sqrt{\gamma^2{\beta}{\bar{\alpha}}{P_1}}\right)^2}$,\label{L24}
\end{enumerate}
where $\alpha\in{[0,1]}$, $\beta\in{[0,1]}$, $\gamma\in{[-1,1]}$, and $\bar{\alpha}\triangleq{1-\alpha}$.
\label{lem:prop}
\end{lemma}
\begin{IEEEproof}
Proof is detailed in Appendix B.
\end{IEEEproof}

Now, to upper bound $R_2$, following from~\eqref{R2G}, we have
\begin{align}
&R_2-\delta_{2,n}\leq\dfrac{1}{n}\sum_{i=1}^{n}{I(U_i,X_{2,i};Y_{1,i}|X_{r_1,i})}\nonumber\\
&={\dfrac{1}{n}\sum_{i=1}^{n}{h(Y_{1,i}|X_{r_1,i})}}-{\dfrac{1}{n}\sum_{i=1}^{n}{h(Y_{1,i}|U_i,X_{2,i},X_{r_1,i})}},
\label{AB}
\end{align}
The second term of \eqref{AB} is $h_1$ which is characterized in Lemma~\ref{lem:h1}.
Define   ${h_2\triangleq \frac{1}{n}\sum_{i=1}^{n}{h(Y_{1,i}|X_{r_1,i})}}$. To establish an upper bound on $R_2$, we need to provide an upper limit on $h_2$. 
Using the definition of $Y_{1,i}$ in~\eqref{GC1} and applying~\eqref{L12} from Lemma~\ref{lem:Gauss}, $h_2$ is upper bounded as 
\begin{align}
h_2&\leq{{\frac{1}{2}\log2\pi{e}\left(\dfrac{1}{n}{\sum_{i=1}^n{E\left[(X_{1,i}+aX_{2,i})^2\right]-S_5+N_1}}\right)}}\nonumber\\
&\leq{{\frac{1}{2}\log2\pi{e}\left(P_1+a^2P_2+2aS_3-S_5+N_1\right)}}\label{B1}\\
&\leq\frac{1}{2}\log2\pi{e}\left(P_1(1-\gamma^2\bar{\alpha}\beta)+a^2{\alpha}P_2\right.\nonumber\\
&\hspace{17mm}+{2a\alpha\gamma}\left.\sqrt{{\beta}P_1P_2}+N_1\right),\label{B}
\end{align}
where~\eqref{B} follows by applying Lemma~\ref{lem:prop}(b) and Lemma~\ref{lem:prop}(d) to~\eqref{B1}. 
Replacing~\eqref{B} and~\eqref{A} in~\eqref{AB} concludes the first upper bound on $R_2$.

Finally, we proceed to establish the second upper bound on $R_2$. Again, by considering the general result in~\eqref{R2G1}, we have
\begin{align}
R_2&-\delta_{2,n}\leq\dfrac{1}{n}\sum_{i=1}^{n}{I(U_i,X_{2,i},X_{r_1,i};Y_{2,i})}\nonumber\\
&={\dfrac{1}{n}\sum_{i=1}^{n}{h(Y_{2,i})}}-{\dfrac{1}{n}\sum_{i=1}^{n}{h(Y_{2,i}|U_i,X_{2,i},X_{r_1,i})}}.\label{CD}
\end{align}
To establish an upper bound on $R_2$, it suffices to establish an upper limit on ${h_3 \triangleq \frac{1}{n}\sum_{i=1}^{n}{h(Y_{2,i})}}$ and a lower limit on ${h_4\triangleq \frac{1}{n}\sum_{i=1}^{n}{h(Y_{2,i}|U_i,X_{2,i},X_{r_1,i})}}$. Following from~\eqref{GC2} and using~\eqref{L12}, $h_3$ can be upper bounded as
\begin{align}
\hspace{-2mm}h_3&\leq{\frac{1}{2}\log2\pi{e}\left(\dfrac{1}{n}\sum_{i=1}^{n}{E[(Y_{1,i}+X_{r_1,i})^2+N_2]}\right)}\nonumber\\
&\leq\frac{1}{2}\log2\pi{e}\left(\dfrac{1}{n}\sum_{i=1}^n{E[X_{1,i}^2]}+a^2\dfrac{1}{n}\sum_{i=1}^n{E[X_{2,i}^2]}\right.\nonumber\\
&\hspace{10mm}+\dfrac{1}{n}\sum_{i=1}^n{E[X_{r_1,i}^2]}\left.+2aS_3+2S_4+N_1+N_2\right)\nonumber\\
&\leq\frac{1}{2}\log2\pi{e}\left(P_1+a^2P_2+P_{r_1}+N_1+N_2\right.\nonumber\\
&\hspace{3mm}\left.+2a\gamma\sqrt{\beta{P_1P_2}}+2\gamma\sqrt{\bar{\alpha}\beta{{P_{r_1}P_1}}}+2a\sqrt{\bar{\alpha}{P_{r_1}P_2}}\right),\label{C}
\end{align}
where~\eqref{C} follows from~\eqref{PowerConst}, Lemma~\ref{lem:prop}(b) and Lemma~\ref{lem:prop}(c).

Next, we establish a lower bound on $h_4$ as 
\vspace{-2mm}
\begin{align}
\hspace{-3mm}h_4&={1}/{n}\sum_{i=1}^{n}h(Y_{2,i}|U_i,X_{2,i},X_{r_1,i})\nonumber\\
&\geq{1}/{2n}\sum_{i=1}^{n}\log\left(2^{2{h(Y_{1,i}|U_i,X_{2,i},X_{r_1,i})}}+2\pi{e}N_2\right)\label{DDD}\\
&\geq{1}/{2}\log\left(2^{2{\dfrac{1}{n}\sum_{i=1}^{n}h(Y_{1,i}|U_i,X_{2,i},X_{r_1,i})}}+2\pi{e}N_2\right)\label{D1}\\
&={1}/{2}\log2\pi{e}\left((1-\gamma^2)P_1+N_1+N_2\right)\label{D}
\end{align}
where~\eqref{DDD} follows using~\eqref{GC2} and considering a generalized entropy power inequality stated in Lemma~\ref{lem:entropy-power} below, \eqref{D1} follows because $\log(2^x+c)$ is a convex function of $x$, and~\eqref{D} follows from~\eqref{A}.
Considering the upper limit in~\eqref{C} and the lower limit in~\eqref{D}, the desired bound on~\eqref{CD} is obtained, leading to the second upper bound on $R_2$.
\begin{lemma}
Consider random variables $X$, $Y$, and $Z$ such that $Y$ and $Z$ are independent, i.e., $I(Y;Z) = 0$, then 
\begin{equation}
2^{2h(X+Y|Z)} \geq 2^{2h(X|Z)} + 2^{2h(Y)}.
\vspace{-4mm}
\end{equation}
\label{lem:entropy-power}
\end{lemma}
\vspace{-3mm}
\begin{IEEEproof}
Proof is detailed in Appendix C.
\end{IEEEproof}

This completes the proof and hence the characterization of the capacity region of the Gaussian degraded CIC-UDC which was open prior to this work.

\vspace{-3mm}
\appendices
\section{Proof of Lemma~\ref{lem:Gauss}}
\vspace{-1mm}
Due to space considerations, $\sum_{i=1}^n(\cdot)$ is shown by $\sum(\cdot)$. Inequality~\eqref{L11} follows as
\begin{align}
&\left|{1}/{n}\sum{E[V_{1,i}V_{2,i}]}\right|=\left|{1}/{n}\sum{E\left[V_{1,i}E\left[V_{2,i}|V_{1,i}\right]\right]}\right|\nonumber\\
&\hspace{5mm}\leq{1}/{n}\sum\sqrt{E[V_{1,i}^2]E[E^2[V_{2,i}|V_{1,i}]]}\nonumber\\
&\hspace{5mm}\leq\sqrt{{1}/{n}\sum{E[V_{1,i}^2]}{1}/{n}\sum{E[E^2[V_{2,i}|V_{1,i}]]}}
=\sqrt{K\times{L}}.\nonumber
\end{align}

Inequality~\eqref{L12} follows from [14, Lemma 1, Part 3].
\hfill $\blacksquare$
\vspace{-4mm}
\section{Proof of Lemma~\ref{lem:prop}}
To prove (a), we first note that $h_1$, defined in Lemma~\ref{lem:h1}, can be bounded as
\begin{align}
h_1&\leq{{1}/{n}\sum{h(X_{1,i}+Z_{1,i}|U_i,X_{2,i},X_{r_1,i})}}\nonumber\\
&\leq{{1}/{n}\sum{h(X_{1,i}+Z_{1,i}|X_{r_1,i})}}\nonumber\\
&\leq{{1}/{2}\log2\pi{e}\left({1}/{n}\sum{E[X_{1,i}^2+Z_{1,i}^2]-S_1}\right)}\label{q1}\\
&\leq{{1}/{2}\log2\pi{e}(P_1-S_1+N_1)}\label{q2},
\end{align}
where~\eqref{q1} follows from~\eqref{L12}. From~\eqref{A} and~\eqref{q2} we have
\begin{equation}
S_1={1}/{n}\sum{E\left[E^2[X_{1,i}|X_{r_1,i}]\right]}\leq{\gamma^2{P_1}}.
\end{equation}
Similarly it can be shown that 
\begin{equation}
S_2={1}/{n}\sum{E\left[E^2[X_{1,i}|X_{2,i}]\right]}\leq{\gamma^2{P_1}}.
\end{equation}
Therefore, there exists some $\beta\in[0,1]$ such that
\begin{equation}
\max\{S_1,S_2\}=\beta\gamma^2{P_1}.\label{SS}
\end{equation}

Next, to show (b), we apply~\eqref{L11} to $S_3$, and then use~\eqref{PowerConst} and~\eqref{SS}. Therefore,

\begin{align}
\left|S_3\right|&\leq{\sqrt{\left({1}/{n}\sum{E[X_{2,i}^2]}\right)\left(S_2\right)}}
\leq{\sqrt{\gamma^2\beta{P_1P_2}}}.
\end{align}

Then, to prove (c), we apply the triangular inequality \cite{c6} to $S_4$, which results in
\begin{align}
\hspace{-3mm}|S_4|&\leq\left|{1}/{n}\sum{E[X_{r_1,i}X_{1,i}]}\right|+a\left|{1}/{n}\sum{E[X_{r_1,i}X_{2,i}]}\right|\nonumber\\
&\leq{\sqrt{\frac{1}{n}\sum{E[X_{r_1,i}^2]}}\left(a\sqrt{\frac{1}{n}\sum{E[X_{2,i}^2]}}+\sqrt{S_1}\right)}\label{W1}\\
&\leq{\sqrt{P_{r_1}}\left(a\sqrt{P_2}+\sqrt{\gamma^2{\beta}{P_1}}\right)}\label{W2},
\end{align}
where~\eqref{W1} follows from~\eqref{L12}, and~\eqref{W2} follows from~\eqref{PowerConst} and~\eqref{SS}. Therefore, there exists some $\bar{\alpha}\in[0,1]$ such that
\begin{equation}
|S_4|=\sqrt{P_{r_1}}\left(a\sqrt{\bar{\alpha}P_2}+\sqrt{\gamma^2{\beta}{\bar{\alpha}}{P_1}}\right).\label{s4}
\end{equation}

Finally, to prove (d), we apply~\eqref{L11} to $S_4$, so
\begin{align}
|S_4|&\leq{\sqrt{S_5\times{1}/{n}\sum{E[X_{r_1,i}^2]}}}\leq{\sqrt{S_5P_{r_1}}}.\label{s5}
\end{align}
Comparing~\eqref{s4} and~\eqref{s5} results in
\begin{equation}
S_5\geq{\left(a\sqrt{\bar{\alpha}P_2}+\sqrt{\gamma^2{\beta}{\bar{\alpha}}{P_1}}\right)^2}.
\end{equation}

\vspace{-4mm}
\section{Proof of Lemma~\ref{lem:entropy-power}}
\vspace{-6mm}
\begin{align}
2h(X+Y|Z) &= 2E[h(X+Y|Z=z)] \nonumber\\
&\geq E \left[ \log \left(2^{2h(X|Z=z)} + 2^{2h(Y)}\right) \right]\label{eq:pow-ent}\\
&\geq  \log \left(2^{2E [h(X|Z=z)]} + 2^{2h(Y)}\right) \label{eq:convex}\\
&= \log \left(2^{2h(X|Z)} + 2^{2h(Y)}\right),
\end{align}
where~\eqref{eq:pow-ent} follows from the entropy power inequality~\cite{c6} and~\eqref{eq:convex} follows from the convexity of $\log (2^x+c)$. \hfill $\blacksquare$

\end{document}